\documentclass[preprint,12pt]{elsarticle}

\usepackage{amssymb}
\usepackage{array}
\usepackage{amsmath}
\usepackage{url}
\usepackage{booktabs} 
\usepackage{xcolor}
\usepackage{braket}
\usepackage{longtable}  
\usepackage{pdflscape}  
\usepackage{physics}
\usepackage{booktabs}

\journal{Nuclear Physics B}

\begin{document}

\begin{frontmatter}



\title{CHIPS-TB: Evaluating Tight-Binding Models For Metals, Semiconductors, and Insulators}


 \author[label1,label2]{In Jun Park}
 \ead{injun.park@nist.gov}


 \author [label1,label3,label4]{Kamal Choudhary}
 \ead{kchoudh2@jhu.edu}
 \affiliation[label1]{organization={National Institute of Standard and Technology},
             city={Gaithersburg},
             state={MD},
             country={USA}}

 \affiliation[label2]{organization={University of Maryland},
             city={College Park},
             state={MD},
             country={USA}}
\affiliation[label3]{Department of Materials Science and Engineering, Whiting School of Engineering, The Johns Hopkins University, Baltimore, MD 21218, USA}
\affiliation[label4]{Department of Electrical and Computer Engineering, Whiting School of Engineering, The Johns Hopkins University, Baltimore, MD 21218, USA}

\begin{abstract}
As semiconductor technologies continue to scale down to the nanoscale, the efficient prediction of material properties becomes increasingly critical. The tight-binding (TB) method is a widely used semi-empirical approach that offers a computationally tractable alternative to Density Functional Theory (DFT) for large-scale electronic structure calculations. However, conventional TB models often suffer from limited transferability and lack standardized benchmarking protocols. In this study, we introduce a computational framework (CHIPS-TB) for evaluating and comparing tight-binding parameterizations across diverse material systems relevant to semiconductor design, focusing on properties such as electronic bandgaps, band structures, and bulk modulus. We assess model parameterizations including Density Functional Tight-Binding (DFTB)-based MatSci, PBC, PTBP, SlaKoNet and TB3PY against \textcolor{black}{OptB88vdW, TBmBJ-DFT and experimental} reference data from the JARVIS-DFT database for 50+ materials pertinent to semiconductor applications. 
The CHIPS-TB code will be made publicly available on GitHub and benchmarks will be available on JARVIS-Leaderboard.
%
\end{abstract}

\end{frontmatter}
\newpage
\section{Introduction}
The continued miniaturization of semiconductor devices has amplified the demand for accurate and efficient predictive modeling of material properties.
While Density Functional Theory (DFT) \cite{sholl2022density} remains a cornerstone of first-principles electronic structure calculations, its high computational cost limits its applicability to large or complex systems, particularly in high-throughput and device-scale simulations \cite{todorov2002tight,luisieR2006atomistic}.

To overcome these limitations, the tight-binding (TB) method has emerged as a widely adopted semi-empirical alternative \cite{slater1954simplified,mehl1996applications,porezag1995construction,koskinen2009density,bannwarth2019gfn2,groth2014kwant,yusufaly2013tight,hourahine2020dftb+,seifert1996calculations,elstner1998self}.
By simplifying the electronic structure problem through the use of localized atomic orbitals and empirical hopping parameters, TB methods offer orders-of-magnitude improvements in computational efficiency compared to DFT. The term tight in tight-binding refers to the approximation that atomic wave function overlap is confined primarily to neighboring atoms. This simplification enables the omission of long-range interactions in the Hamiltonian, significantly reducing computational complexity. The TB method is a semi-empirical computational approach capable of calculating electronic band structures while accounting for external influences such as strain, electric fields, and optical radiation on the material system \cite{kohleR2005density,pu2004combining,vasudevan2019materials,schledeR2019dft,schmidt2019recent,dftb_pt1, dftb_pt2, GFNxTB, vdw}.
This has enabled their use in applications ranging from nanodevice modeling to materials discovery.

The TB approach is primarily employed to compute the band structure and single-particle Bloch states of materials. Parameters for the TB Hamiltonian, typically expressed in real-space, are derived from first-principles calculations to ensure physical accuracy. This parameterized Hamiltonian can then be applied to simulate very large systems, including both periodic and non-periodic structures, at a significantly reduced computational cost. Constructing a TB Hamiltonian involves developing a model that accurately reproduces the band energies obtained from reference first-principles methods.


Several variants of the TB method have been developed to improve accuracy, transferability, and computational efficiency across diverse material systems. At the core of many of these approaches is the Slater-Koster (SK) formalism, which parameterizes hopping integrals based on orbital symmetries and directional cosines. This foundational model provides a versatile framework that underpins much of modern TB methodology. Among the more advanced variants, Wannier-function-based TB models derive parameters directly from first-principles calculations, offering a high degree of accuracy while preserving computational efficiency \cite{marzari2012maximally}. The Density Functional Tight-Binding (DFTB) method extends traditional TB by incorporating self-consistent charge corrections and repulsive potentials, making it a popular choice for complex systems \cite{elstner1998self,kohleR2005density}.

Further refinements include the three-body tight-binding (TB3PY) model, which introduces additional interaction terms to improve band structure predictions across a wider range of materials \cite{garrity2023fast}. Semiempirical methods such as Parametric Method 6 with D3 Dispersion, H4 Hydrogen-Bonding, and Halogen-Bonding Corrections (PM6-D3H4X) \cite{stewart2007optimization}, Third-order Density Functional based Tight-Binding method augmented with Grimme’s D3 dispersion correction using Becke-Johnson damping (DFTB3-D3(BJ)), and Geometry, Frequency, Noncovalent, version 2 - extended Tight-Binding (GFN2-xTB) \cite{bannwarth2019gfn2} enhance standard TB approaches by integrating dispersion corrections and multipole electrostatics, extending their applicability to molecular and condensed-phase systems. More recently, the Parameterized Tight-Binding (PTB) model has been proposed to generate transferable Hamiltonians based on high-throughput data fitting, aiming to bridge the gap between accuracy and generalizability \cite{cui2024obtaining}. \textcolor{black}{Another advanced approach is SlaKoNet, a neural network framework based on the Slater-Koster tight-binding formalism \cite{choudhary2025slakonet}. SlaKoNet aims to optimize TB parameters by training on higher-level DFT data, such as meta-GGA bandgaps, using PyTorch-based neural networks for improved scalability and accuracy across diverse systems.} Collectively, these models represent a spectrum of trade-offs between physical rigor and computational speed, highlighting the need for systematic benchmarking across material classes to guide the development of reliable TB parameterizations.

Despite the promise of TB methods, a major challenge remains: transferability.
TB models, being parameter-dependent, often fail to maintain accuracy across chemically and structurally diverse systems.
This highlights the need for systematic benchmarking frameworks that can quantitatively assess their predictive capabilities relative to trusted reference data. Some of the previous works pertaining to TB benchmarking include works by Gruden et al. for reaction energetics \cite{gruden2017benchmarking}, Oliveira et al. \cite{oliveira2016benchmarking} for small clusters.
\textcolor{black}{While previous benchmarking efforts (e.g., CHIPS-FF) have focused on evaluating interatomic force fields \cite{wines2025chips}, the unique characteristics of electronic structure prediction with tight-binding models necessitate a specialized framework. Evaluating band structures, bandgaps, and electronic states requires distinct metrics and analyses to understand how TB approximations impact charge distribution, electron-hole interactions, and band dispersion. This work, CHIPS-TB, provides a unique contribution by focusing specifically on these electronic properties and offering insights into the underlying physics captured by different TB formalisms.}

In this work, we present a computational benchmarking framework for evaluating TB models using reference data from the JARVIS-DFT database \cite{choudhary2020joint,wines2023recent,choudhary2025jarvis}.
We focus on two representative TB approaches, DFTB and TB3PY\textcolor{black}{, along with SlaKoNet,} and benchmark their performance across \textcolor{black}{50+} materials relevant to semiconductor applications.
By comparing key electronic structure metrics such as bandgaps and eigenvalue spectra, we assess the strengths and limitations of each model, with the goal of guiding future development of transferable and efficient TB parameterizations.


\section{Methodology}

\begin{figure}[th]
\centering
\includegraphics[width=1\linewidth]{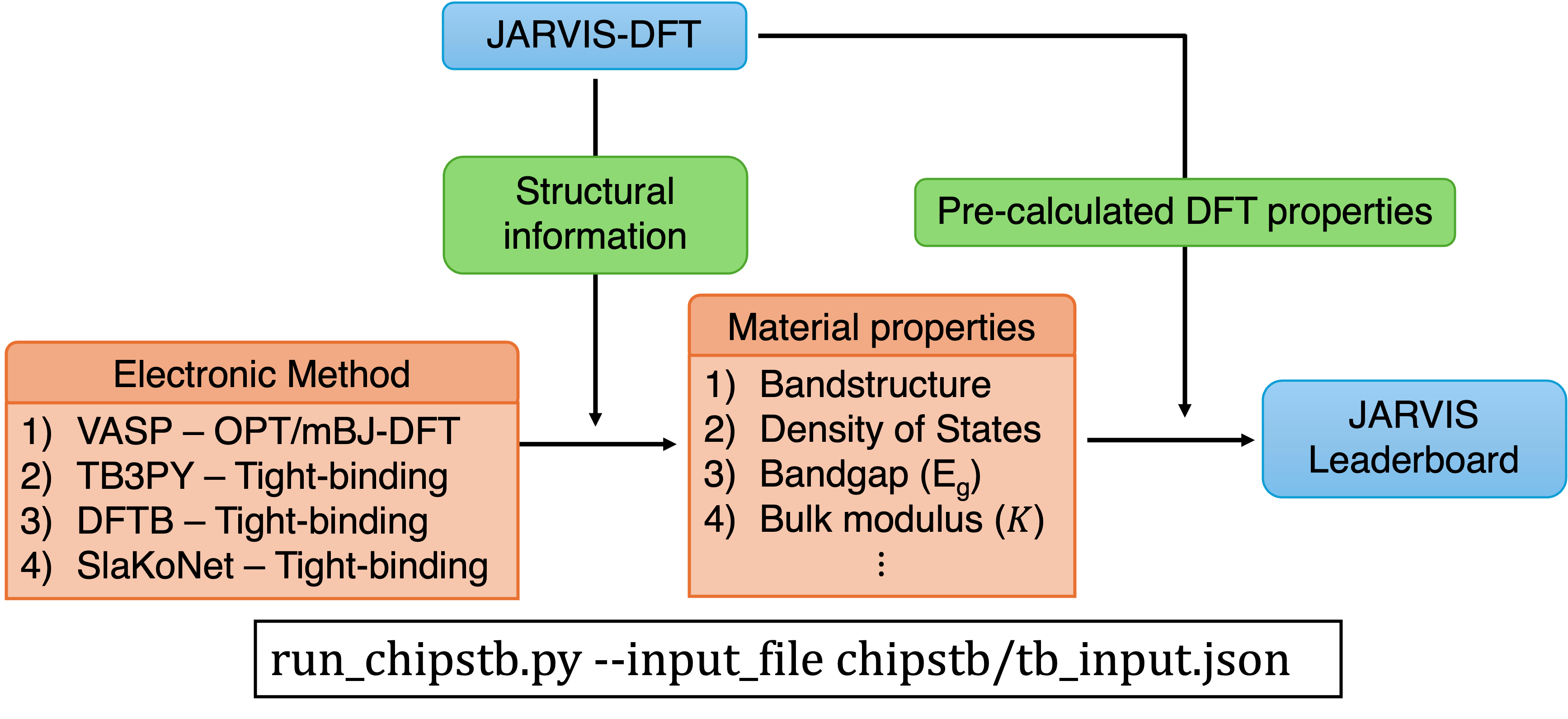}

\caption{Workflow for the evaluation of tight-binding models. Structural information and pre-calculated DFT properties are obtained from the JARVIS-DFT database. The structural data serves as input to various electronic structure methods-including VASP \textcolor{black}{(OptB88vdW (OPT)-DFT)}, TB3PY (tight-binding), and DFTB (tight-binding) to compute key material properties such as bandstructure, density of states, bandgap (E$_g$), and bulk modulus ($K$). These computed properties are compared against DFT references using the $run\_chipstb.py$ script with a configuration file (e.g., $tb\_input.json$). The resulting performance metrics are reported on the JARVIS Leaderboard to evaluate the accuracy of different tight-binding models.}\label{workflow}
\end{figure}

We developed a computational framework for benchmarking tight-binding (TB) models against reference data from the JARVIS-DFT database, which includes \textcolor{black}{OptB88vdW (OPT)-DFT}~\cite{klimevs2009chemical} and \textcolor{black}{TB-mBJ}~\cite{tran2009accurate} calculations performed using the Vienna Ab-initio Simulation Package (VASP)~\cite{kresse1996efficiency,kresse1996efficient}, the JARVIS-Tools package, as well as available experimental data. The methodology consists of three main steps: (1) extracting structural information from the JARVIS-DFT repository, (2) computing electronic and mechanical properties using different TB models, and (3) quantitatively comparing the results with DFT reference values as shown in Figure 1.

\textcolor{black}{The TB method approximates the electronic structure of solids by expanding the crystal wave function ($\psi_i$) in a basis of a linear combination of atomic orbitals (LCAO).}

\textcolor{black}{
\begin{equation}
\psi_{i}(\mathbf{r}) = \sum_{u}c_{iu}\phi_u(\mathbf{r})
\end{equation}
}
\textcolor{black}{where $c_{iu}$ denote the expansion coefficients, and $\phi_u(\mathbf{r})$ describes the $u$-th atomic orbital. The equation leads to the generalized eigenvalue problem as follows:}

\textcolor{black}{
\begin{equation}
\sum_{v}H_{uv}c_{iv} = \epsilon_{i}\sum_{v} S_{uv} c_{iv},
\end{equation}
}

\textcolor{black}{where $H_{uv} = \Braket{\phi_u | \hat{H} | \phi_v}$ and $S_{uv} = \Braket{\phi_u | \phi_v}$ are the Hamiltonian and overlap matrices, respectively, in the basis of atomic orbitals.}

\textcolor{black}{The tight-binding (TB) method approximates the electronic structure of solids by expressing the Hamiltonian in a basis of localized atomic orbitals. Unlike traditional orthogonal tight-binding approaches, which typically use simple hopping integrals as described by generic Hamiltonian forms, the Density Functional Tight-Binding (DFTB) \cite{elstner1998self,kohleR2005density} and Three-Body Tight-Binding (TB3PY) \cite{garrity2023fast} methods employed here utilize more sophisticated and physically grounded formalisms. Both methods use non-orthogonal basis sets and are parameterized through distinct approaches.}

\textcolor{black}{DFTB is an approximate Kohn-Sham Density Functional Theory (DFT) method, derived from a second-order expansion of the DFT total energy, aiming to reproduce full DFT results with significantly reduced computational cost \cite{elstner1998self,kohleR2005density}. The total energy expression in DFTB is obtained by expanding the Kohn-Sham Hamiltonian around a superposition of atomic densities. Crucially, the Hamiltonian and overlap matrices are pretabulated through full self-consistent field (SCF) DFT calculations of atomic dimers. This means DFTB parameters are not directly fitted to band structures in an ad hoc empirical manner, but rather derived from ab initio calculations. The accuracy of DFTB heavily depends on choices made during the tabulation process, such as the specific density and wavefunction confining radii used for each atomic species. These radii significantly influence the basis set and, consequently, the agreement with electronic structure metrics. For example, specific choices can lead to highly accurate electronic band structures for materials like silicon \cite{markov2015atomic}. DFTB incorporates self-consistent charge corrections and short-range repulsive potentials to account for charge redistribution and core-core repulsion. In this study, among the available DFTB-style parameterizations (MatSci\cite{guimaraes2007imogolite}, PBC\cite{rauls1999stoichiometric}, PTBP\cite{cui2024obtaining} and SlaKoNet\cite{choudhary2025slakonet}).}

\textcolor{black}{TB3PY is a tight-binding method developed to provide fast and accurate predictions of material properties, particularly for the periodic table \cite{garrity2023fast}. Similar to DFTB, TB3PY also utilizes a non-orthogonal basis set and includes self-consistent charge equilibration steps to account for charge transfer effects. The defining characteristic of TB3PY is the explicit inclusion of angular-dependent three-body terms in the Hamiltonian. These terms are designed to improve the description of covalent bonding and local environment effects, which are critical for accurately capturing band dispersion and curvature in many materials, including some covalent semiconductors and metallic states. The Hamiltonian and overlap matrices in TB3PY are expressed in terms of polynomial expansions whose coefficients are determined through fitting to selected ab initio datasets. This makes TB3PY a more empirical method than DFTB+, as its parameters are explicitly optimized against reference data. The repulsive energy contribution in TB3PY is accounted for as an eigenvalue shift, rather than an external function as in some other TB approaches. This method's formulation helps in capturing more complex interactions, often leading to improved accuracy in certain systems.}

\textcolor{black}{SlaKoNet \cite{choudhary2025slakonet} is a neural network framework based on the Slater-Koster tight-binding optimization model. It was inspired by methods like TB3PY but differentiates itself by being trained on higher-level DFT data, specifically meta-GGA bandgaps (like those from mBJ-DFT), rather than solely semi-local bandgaps. Instead of conventional least-squares fitting, SlaKoNet employs a PyTorch-based neural network for optimizing model parameters, which has demonstrated promising scalability for both CPU and GPU architectures. In this study, we consider SlaKoNet's performance for electronic properties.}

\textcolor{black}{The OPT-DFT and TBmBJ results from the JARVIS-DFT database are utilized as our reference data. This approach ensures internal consistency for benchmarking DFTB and TB3PY models, as these tight-binding methods are commonly parameterized or fitted to PBE-level information. While acknowledging that PBE is known to systematically underestimate bandgaps compared to experimental values or higher-level ab initio methods like GW \cite{aryasetiawan1998gw} and modified Becke-Johnson (mBJ) potentials \cite{tran2009accurate}, using PBE as a consistent baseline allows for a direct evaluation of the TB models' fidelity in reproducing OPT-DFT-like electronic structures. TB parameters are fit to these semi-local DFT methods, and thus direct comparison to experiment would not be an accurate measure of the TB model's faithfulness to its training data.}

\textcolor{black}{Long-Range Corrected DFTB (LC-DFTB) has been developed and parameterized over the last decade to address known limitations of semi-local DFTB, particularly for band edges and charge-transfer physics in semiconductors and insulators \cite{lutskeR2015implementation}. LC-DFTB variants (e.g., LC-DFTB2 \cite{vuong2018parametrization} and TD-LC-DFTB \cite{sokolov2021analytical}) offer improved accuracy in these challenging areas. Future work will explore the inclusion of LC-DFTB parameterizations as they become more broadly available and validated across a wide material spectrum.}


The benchmarking analysis was performed on a diverse set of \textcolor{black}{50+} materials selected from the JARVIS-DFT database to represent a broad range of bonding types and electronic characteristics relevant to semiconductor and electronic materials. The test set includes elemental systems such as silicon (JVASP-1002), carbon (JVASP-91), aluminum (JVASP-816), copper (JVASP-867), gold (JVASP-825), and titanium (JVASP-1029), covering both semiconducting and metallic behavior. Binary compounds include silicon carbide (JVASP-8118, 8158, 107), silicon dioxide (JVASP-41, 34674), gallium arsenide (JVASP-1174), aluminum phosphide (JVASP-1327), and zinc oxide (JVASP-1195), offering a mixture of covalent, ionic, and polar materials. This selection ensures that the benchmarking captures a range of structural motifs (zincblende, rock-salt, wurtzite, and rutile types), bonding chemistries, and bandgap magnitudes, thus providing a comprehensive assessment of TB model performance across technologically relevant materials.

The calculations were performed along high-symmetry $\mathbf{k}$-paths defined in the Brillouin zone. To assess model performance, we employed several evaluation metrics. First, the maximum deviation in eigenvalues between OPT-DFT and TB across all $\mathbf{k}$-points and bands was computed as:

\begin{equation}
\textcolor{black}{\Delta_{\text{max}} = \max_{n, \mathbf{k}} \left| E^{\text{OPT-DFT}}_{n\mathbf{k}} - E^{\text{TB}}_{n\mathbf{k}} \right|.}
\end{equation}

Second, we evaluated the absolute error in the predicted bandgap:

\begin{equation}
\textcolor{black}{\Delta E_g = \left| E^{\text{OPT-DFT}}_g - E^{\text{TB}}_g \right|.}
\end{equation}

Finally, the mean absolute error (MAE) in bandgap across all tested materials was calculated using:

\begin{equation}
\textcolor{black}{\text{MAE}_{E_g} = \frac{1}{N} \sum_{i=1}^{N} \left| E^{\text{OPT-DFT}}_{g,i} - E^{\text{TB}}_{g,i} \right|,}
\end{equation}

where $N$ is the number of materials in the test set. These metrics were used to quantitatively assess the fidelity of TB models in reproducing \textcolor{black}{OPT-DFT-calculated} band structures and electronic gaps. The results of the work will also be made available on the JARVIS-Leaderboard \cite{choudhary2024jarvis} benchmarking platform to enhance transparency and reproducibility. 




\begin{figure*}[htbp]
\centering
\includegraphics[width=\textwidth]{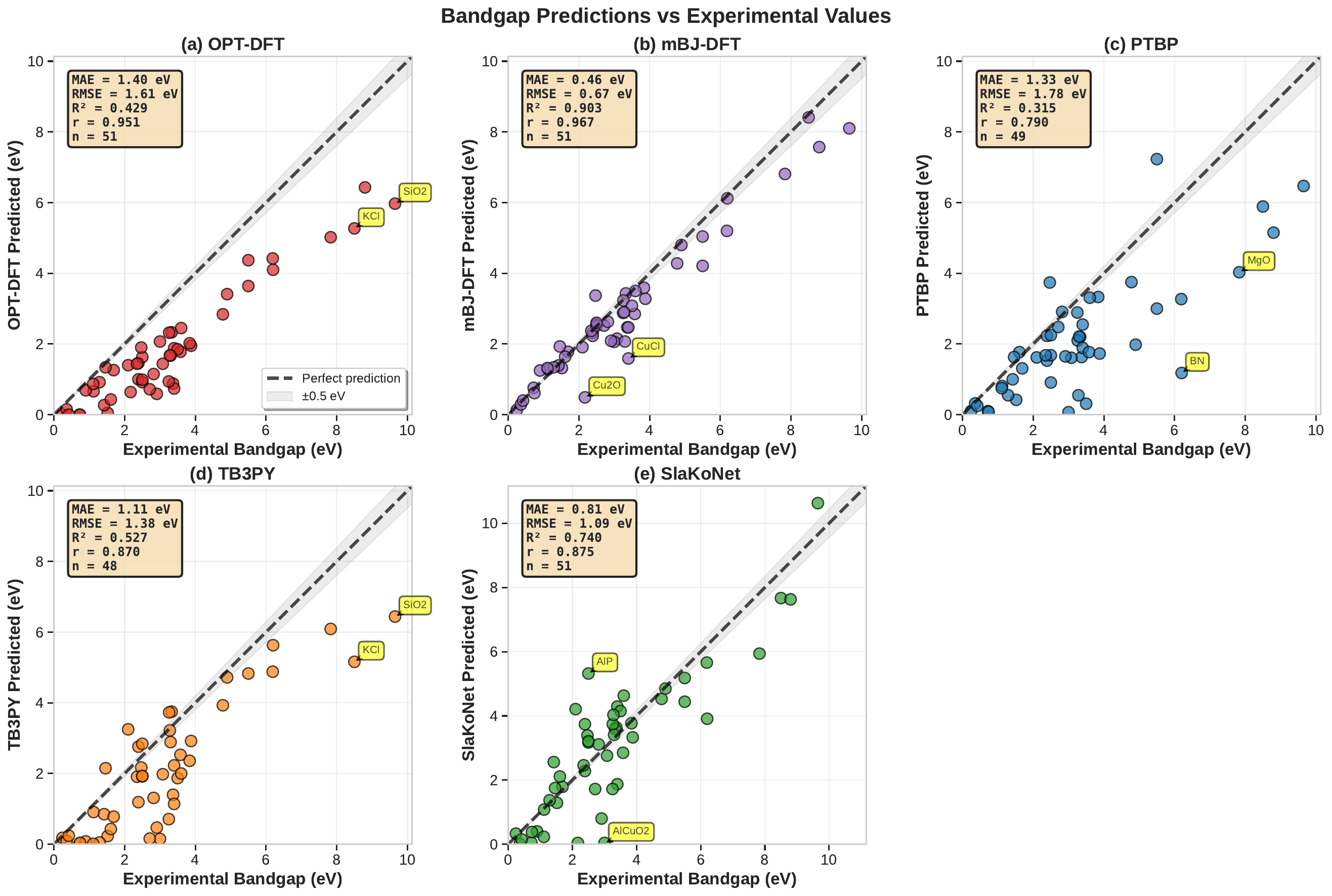}
\caption{\textbf{Bandgap predictions versus experimental values.} Parity plots for (a) OPT-DFT, (b) mBJ-DFT, (c) PTBP, (d) TB3PY, and (e) SlaKoNet methods. The dashed line shows perfect prediction; gray bands indicate ±0.5 eV error. mBJ-DFT achieves the best agreement (MAE: 0.46 eV), followed by SlaKoNet (MAE: 0.81 eV). Among TB methods, TB3PY (MAE: 1.11 eV) outperforms PTBP (MAE: 1.33 eV). Here, MAE represents mean absolute error, RMSE root mean square error, R$^2$ the coefficient of determination, r the Pearson correlation coefficient, and n the number of data points.}
\label{fig:bandgap_vs_exp}
\end{figure*}

\begin{figure}[htbp]
\centering
\includegraphics[width=0.8\textwidth]{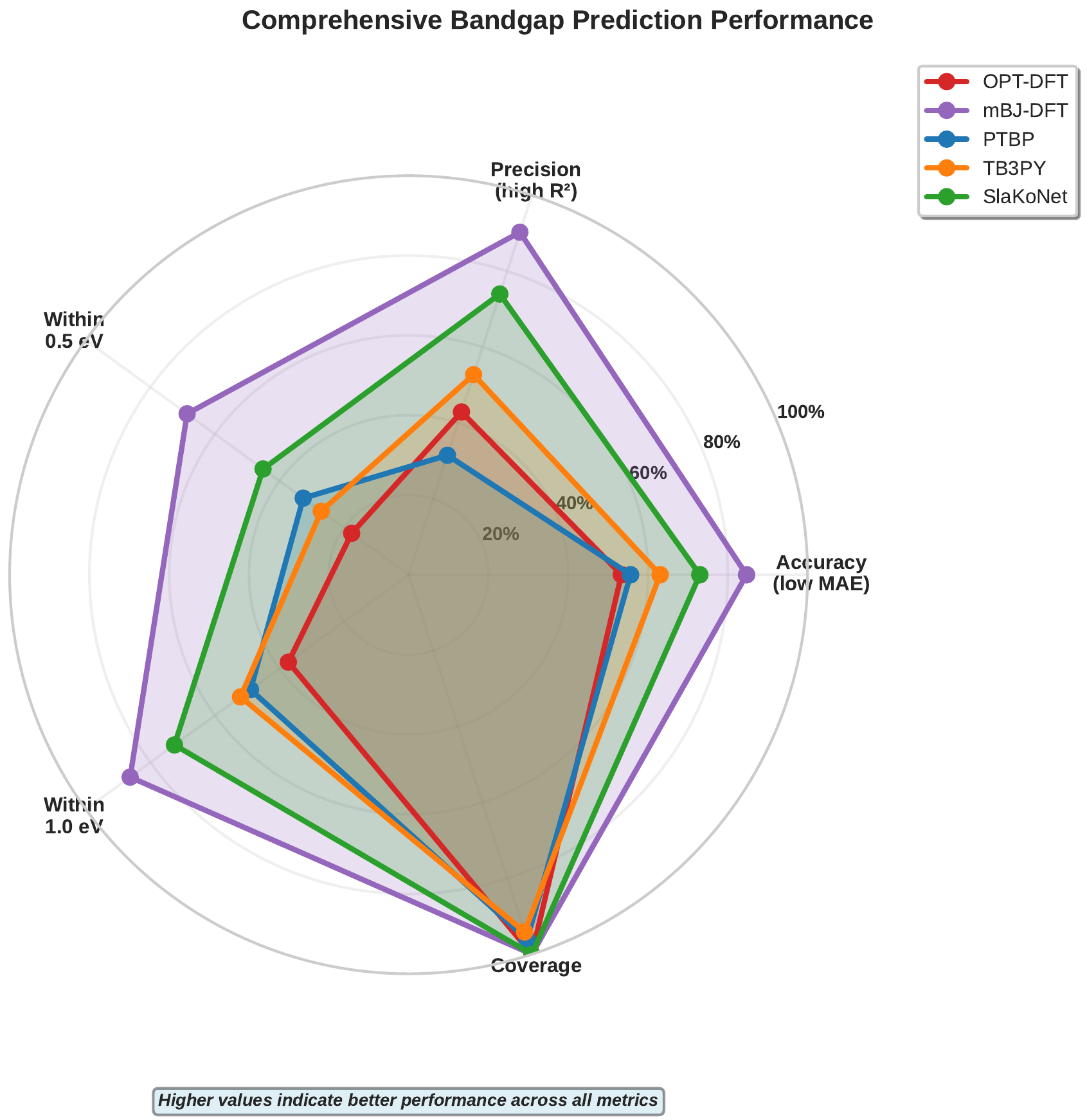}
\caption{\textbf{Comprehensive performance metrics for bandgap predictions.} Radar chart comparing accuracy, precision (R2), percentages within 0.5/1.0 eV thresholds, and data coverage. mBJ-DFT shows best overall performance; SlaKoNet leads among TB methods for experimental validation.}
\label{fig:radar_performance}
\end{figure}

\section{Results and discussion}

To assess generalizability, we extended the bandgap comparison to a set of \textcolor{black}{50+} materials with non-zero bandgaps. Table~\ref{bandgap_comparison} summarizes the \textcolor{black}{predicted bandgaps from the TB models, alongside OPT-DFT, experimental, and mBJ-DFT reference values for contextual comparison\cite{choudhary2018computational}.}  \textcolor{black}{The SlaKoNet model shows competitive performance with an MAE of 1.46 eV against OPT-DFT, and performs exceptionally well against experimental bandgaps with an MAE of 0.46 eV. This highlights SlaKoNet's training on high-accuracy bandgaps.}The MAE for DFTB-PTBP is 1.33~eV, while TB3PY achieves a lower MAE of 1.11~eVvagainst experimnetal data.

\textcolor{black}{The superior performance of TB3PY and SlaKoNet for bandgaps against OPT-DFT, experimental, and mBJ-DFT reference data is attributed to their parameterization strategies. TB3PY's inclusion of three-body terms allows for a more flexible description of chemical bonding, and SlaKoNet's neural network approach trained on mBJ-DFT data is explicitly designed to capture higher-accuracy bandgaps. Conversely, DFTB's PTBP parameterization, despite its semi-local DFT approximations, demonstrates a higher MAE in bandgap prediction across this larger and more diverse test set. This highlights that the choice of parameterization within a given formalism is crucial for specific properties, and that methods trained on or incorporating features relevant to higher-accuracy bandgaps (like mBJ-DFT data for SlaKoNet) can achieve superior predictive power.} 

\textcolor{black}{Figure~\ref{fig:bandgap_vs_exp} compares bandgap predictions against experimental values for 48-51 materials. mBJ-DFT achieves the best agreement (MAE: 0.46 eV, R$^2$: 0.903), while SlaKoNet leads among tight-binding methods (MAE: 0.81 eV, R$^2$: 0.740). TB3PY outperforms PTBP (1.11 vs 1.33 eV MAE), and OPT-DFT shows expected semi-local functional underestimation (MAE: 1.40 eV). Outliers exceeding 1.5 eV error occur primarily for wide-gap insulators.}

\textcolor{black}{Figure~\ref{fig:radar_performance} provides multi-dimensional performance analysis across accuracy, precision, error thresholds, and coverage. mBJ-DFT demonstrates the most balanced high performance, while SlaKoNet emerges as the best tight-binding method for experimental validation, achieving >60\% of materials within 1.0 eV error tolerance.}

\begin{figure}[th!]
\centering
\includegraphics[width=1\linewidth]{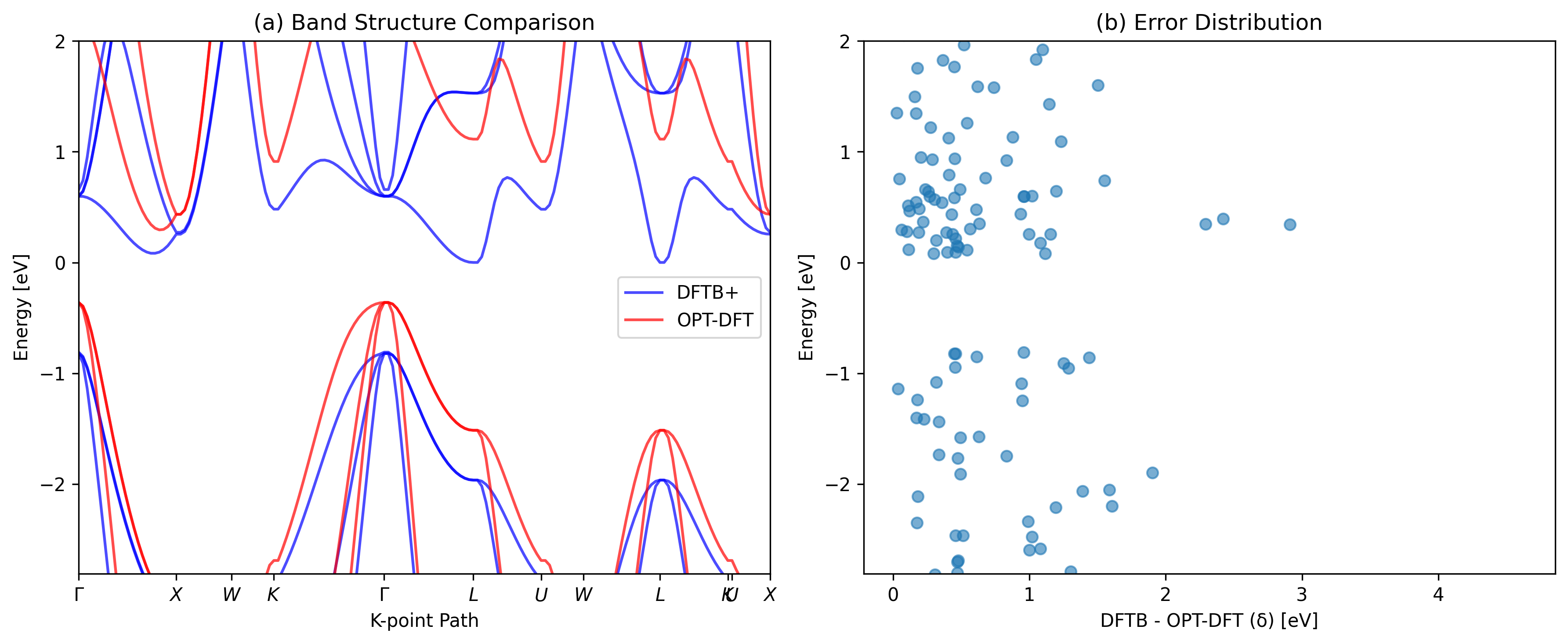}
\includegraphics[width=1\linewidth]{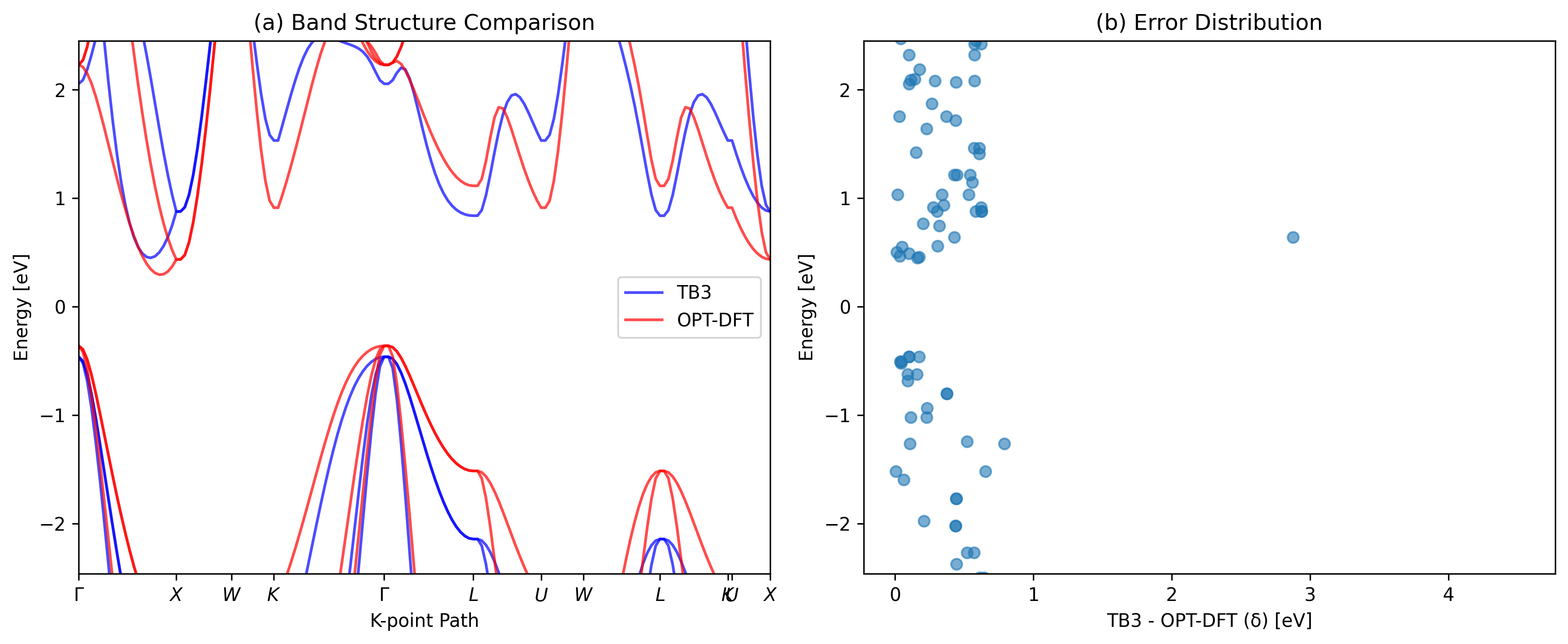}
\caption{Comparison of \textcolor{black}{OPT-DFT} and TB bandstructures for Si. The left panels, (a) and (c), display the full band structures calculated using the DFTB and TB3PY models, respectively. The right panels, (b) and (d), show the pointwise differences between the \textcolor{black}{OPT-DFT} and TB predictions.}
\label{sicompare}
\end{figure}

Next, Figure~\ref{sicompare} presents a comparison of the band structure of silicon in the diamond crystal structure (JVASP-1002), as obtained from \textcolor{black}{OPT-DFT} (red curves) and two TB models (blue curves). The left panels, (a) and (c), display the full band structures calculated using the DFTB and TB3PY models, respectively. The right panels, (b) and (d), show the pointwise differences between the \textcolor{black}{OPT-DFT} and TB predictions, defined as \textcolor{black}{$E^{\text{OPT-DFT}}_{n\mathbf{k}} - E^{\text{TB}}_{n\mathbf{k}}$,} where $n$ is the band index and $\mathbf{k}$ denotes the wavevector along the high-symmetry Brillouin zone path \cite{garrity2021database}.

\begin{landscape}
\footnotesize
\setlength{\tabcolsep}{2pt}
\begin{longtable}{@{}l>{\centering\arraybackslash}p{0.7cm}>{\centering\arraybackslash}p{1.3cm}*{8}{>{\centering\arraybackslash}p{0.75cm}}@{}}
    \caption{Comparison of bandgaps (in eV) from TB methods with DFT and experimental data. Dashes indicate unavailable data.}
    \label{bandgap_comparison} \\
    \toprule
    Mat. & JV\# & Spg. & Exp & mat & PBC & PTB & TB3 & OPT & mBJ & SK \\
    \midrule
    \endfirsthead

    \caption*{Table \thetable{} - Continued from previous page} \\
    \toprule
    Mat. & JV\# & Spg. & Exp & mat & PBC & PTB & TB3 & OPT & mBJ & SK \\
    \midrule
    \endhead

    \midrule
    \multicolumn{11}{r@{}}{\textit{Continued on next page}} \\
    \endfoot

    \bottomrule
    \endlastfoot

    GaAs & 1174 & F$\bar{4}$3m & 1.52 & - & - & 0.42 & 0.23 & 0.05 & 1.32 & 1.29  \\
    Si & 1002 & Fd$\bar{3}$m & 1.12 & 1.82 & 1.23 & 0.81 & 0.91 & 0.66 & 1.28 & 1.08  \\
    ZnO & 1195 & P6$_3$mc & 3.37 & - & - & 1.63 & 1.4 & 0.88 & 2.47 & 3.64  \\
    SiC & 8118 & P6$_3$mc & 3.33 & 2.93 & 5.11 & 2.2 & 3.75 & 2.33 & 3.43 & 3.59  \\
    SiC & 8158 & F$\bar{4}$3m & 2.39 & 2.18 & 5.64 & 1.53 & 2.76 & 1.46 & 2.31 & 3.74  \\
    SiC & 107 & P6$_3$mc & 3.26 & 2.78 & 5.79 & 2.1 & 3.73 & 2.32 & 3.23 & 3.74  \\
    AlP & 1327 & F$\bar{4}$3m & 2.5 & 2.25 & - & 1.68 & 2.84 & 1.63 & 2.56 & 5.32  \\
    C & 91 & Fd$\bar{3}$m & 5.5 & 7.69 & 6.87 & 7.23 & 4.83 & 4.37 & 5.04 & 5.18  \\
    SiO$_2$ & 41 & P3$_2$21 & 9.65 & 8.76 & 9.18 & 6.47 & 6.44 & 5.97 & 8.1 & 10.63  \\
    SiO$_2$ & 34674 & C222$_1$ & - & 8.11 & 9.46 & 6.92 & 6.85 & 5.67 & 8.02 & 10.62  \\
    TiO$_2$ & 104 & I4$_1$/amd & 3.4 & 3.27 & - & 2.55 & 2.23 & 1.88 & 2.47 & 4.29  \\
    ZrO$_2$ & 113 & P2$_1$/c & 5.5 & - & - & 3 & - & 3.64 & 4.21 & 4.44  \\
    KCl & 1145 & Fm$\bar{3}$m & 8.5 & - & - & 5.89 & 5.16 & 5.27 & 8.41 & 7.67  \\
    MgO & 116 & Fm$\bar{3}$m & 7.83 & - & - & 4.03 & 6.09 & 5.02 & 6.81 & 5.94  \\
    InN & 1180 & P6$_3$mc & 0.72 & - & - & 0.1 & 0.01 & 0 & 0.76 & 0.05  \\
    InP & 1183 & F$\bar{4}$3m & 1.42 & - & - & 1 & 0.85 & 0.27 & 1.39 & 2.56  \\
    InSb & 1189 & F$\bar{4}$3m & 0.24 & - & - & 0.1 & 0.18 & 0 & 0.13 & 0.33  \\
    ZnTe & 1198 & F$\bar{4}$3m & 2.39 & - & - & 2.23 & 1.19 & 1 & 2.23 & 2.28  \\
    CuCl & 1201 & F$\bar{4}$3m & 3.4 & - & - & 1.89 & 1.14 & 0.74 & 1.59 & 1.87  \\
    Cu$_2$O & 1216 & Pn$\bar{3}$m & 2.17 & 0.75 & - & - & - & 0.64 & 0.49 & 0.04  \\
    BaTe & 1267 & Fm$\bar{3}$m & 3.08 & - & - & 1.61 & 1.98 & 1.44 & 2.15 & 2.76  \\
    BaSe & 1294 & Fm$\bar{3}$m & 3.58 & - & - & 1.77 & 2.53 & 1.78 & 2.85 & 2.85  \\
    MgS & 1300 & Fm$\bar{3}$m & 4.78 & - & - & 3.75 & 3.93 & 2.84 & 4.28 & 4.53  \\
    BP & 1312 & F$\bar{4}$3m & 2.1 & - & - & 1.62 & 3.25 & 1.4 & 1.91 & 4.21  \\
    BaS & 1315 & Fm$\bar{3}$m & 3.88 & - & - & 1.73 & 2.92 & 1.95 & 3.28 & 3.33  \\
    GaP & 1393 & F$\bar{4}$3m & 2.35 & - & - & 1.68 & 1.91 & 1.44 & 2.37 & 2.46  \\
    AlSb & 1408 & F$\bar{4}$3m & 1.69 & - & - & 1.31 & 0.78 & 1.26 & 1.78 & 1.79  \\
    AlCuO$_2$ & 1453 & R$\bar{3}$m & 3 & 0.92 & - & 0.07 & 0.15 & 2.07 & 2.06 & 0.04  \\
    BN & 17 & P6$_3$/mmc & 6.2 & 3.68 & - & 1.18 & 5.63 & 4.1 & 6.12 & 3.91  \\
    ZnS & 1702 & F$\bar{4}$3m & 3.84 & - & - & 3.33 & 2.36 & 2.02 & 3.59 & 3.77  \\
    AgCl & 1954 & Fm$\bar{3}$m & 3.25 & - & - & 2.89 & 0.71 & 0.94 & 2.88 & 1.72  \\
    CdTe & 23 & F$\bar{4}$3m & 1.61 & - & - & 1.77 & 0.43 & 0.43 & 1.64 & 2.11  \\
    SnSe & 299 & Pnma & 0.9 & - & - & - & 0.08 & 0.69 & 1.25 & 0.4  \\
    GaN & 30 & P6$_3$mc & 3.5 & - & - & 0.31 & 1.87 & 1.85 & 3.08 & 4.15  \\
    Al$_2$O$_3$ & 32 & R$\bar{3}$c & 8.8 & 9.79 & - & 5.15 & - & 6.43 & 7.57 & 7.63  \\
    AlN & 39 & P6$_3$mc & 6.19 & - & - & 3.27 & 4.88 & 4.42 & 5.2 & 5.66  \\
    TiO$_2$ & 5 & P4$_2$/mnm & 3.3 & 3.26 & - & 2.21 & 2.89 & 1.67 & 2.07 & 3.41  \\
    MoS$_2$ & 54 & P6$_3$/mmc & 1.29 & - & - & 0.55 & 0.05 & 0.92 & 1.34 & 1.37  \\
    MoSe$_2$ & 57 & P6$_3$/mmc & 1.11 & - & - & 0.75 & 0.01 & 0.87 & 1.32 & 0.23  \\
    BAs & 7630 & F$\bar{4}$3m & 1.46 & - & - & 1.63 & 2.15 & 1.34 & 1.93 & 1.75  \\
    MgSe & 7678 & Fm$\bar{3}$m & 2.47 & - & - & 3.74 & 2.17 & 1.9 & 3.37 & 3.39  \\
    MgTe & 7762 & F$\bar{4}$3m & 3.6 & - & - & 3.31 & 2 & 2.45 & 3.5 & 4.63  \\
    AlN & 7844 & F$\bar{4}$3m & 4.9 & - & - & 1.98 & 4.72 & 3.41 & 4.8 & 4.85  \\
    SnTe & 7860 & Fm$\bar{3}$m & 0.36 & - & - & 0.32 & 0.09 & 0.15 & 0.29 & 0.01  \\
    CdS & 8003 & F$\bar{4}$3m & 2.5 & - & - & 2.25 & 1.92 & 0.92 & 2.52 & 3.17  \\
    GaN & 8169 & F$\bar{4}$3m & 3.28 & - & - & 0.55 & 3.22 & 1.68 & 2.9 & 4.03  \\
    AgI & 8566 & Fm$\bar{3}$m & 2.91 & - & - & 1.65 & 0.47 & 0.59 & 2.09 & 0.8  \\
    AgBr & 8583 & Fm$\bar{3}$m & 2.71 & - & - & 2.48 & 0.16 & 0.72 & 2.52 & 1.72  \\
    Ge & 890 & Fd$\bar{3}$m & 0.74 & - & - & 0.08 & 0.04 & 0 & 0.61 & 0.38  \\
    CdS & 95 & P6$_3$mc & 2.5 & - & - & 0.91 & 1.93 & 0.99 & 2.6 & 3.21  \\
    ZnSe & 96 & F$\bar{4}$3m & 2.82 & - & - & 2.91 & 1.31 & 1.15 & 2.63 & 3.11  \\
    InAs & 97 & F$\bar{4}$3m & 0.42 & - & - & 0.25 & 0.24 & 0 & 0.4 & 0.15 \\

\midrule
    MAE$_{\text{OPT}}$ & - & - & - & 1.44 & 2.93 & 0.82 & 0.67 & - & 1.07 & 1.46\\
    \midrule
    MAE$_{\text{mBJ}}$ & - & - & - & 0.96 & 1.71 & 0.99 & 0.86 & 1.07 & - & 0.76\\
    \midrule
    MAE$_{\text{Exp}}$ & - & - & - & 0.95 & 1.58 & 1.33 & 1.11 & 1.40 & 0.46 & 0.81\\

\end{longtable}
\end{landscape}

From Figure~\ref{sicompare}, it is evident that the largest discrepancies between \textcolor{black}{OPT-DFT} and TB predictions occur in the conduction bands, where both DFTB and TB3PY show significant deviations from the \textcolor{black}{OPT-DFT} reference. The values of $\text{Diff}_{\text{max}}$ for DFTB and TB3PY are 4.63~eV and 4.54~eV, respectively. 

The predicted bandgaps for silicon further illustrate the similarity in performance. Both DFTB and TB3PY underestimate the bandgap at 0.81~eV and 0.91~eV, respectively, and the \textcolor{black}{JARVIS-OPT-DFT} reference value is 0.656~eV. \textcolor{black}{For contextual comparison, the experimental gap for silicon is approximately 1.12~eV, and the mBJ-DFT value is 1.28~eV. The observed underestimation in both TB models relative to both experiment and mBJ-DFT is a direct consequence of their parameterization to semi-local DFT, which is known to underestimate bandgaps. The observed underestimation in both TB models relative to experiment is a direct consequence of their parameterization to semi-local DFT.} 

\begin{landscape}
\begin{longtable}{@{}lcccccc@{}}
    \caption{Comparison of maximum differences (Max$_{\text{diff}}$, in eV) between TB predictions and \textcolor{black}{OPT-DFT} reference data from the JARVIS database. If data is not available, it's shown by a dash indicating the corresponding parameters do not exist.}
    \setlength{\LTcapwidth}{0.85\textwidth}
    \label{maxdiff_comparison} \\
    \toprule
    Mat. & JV\# & Spg. & Max$_{\text{matsci}}$ & Max$_{\text{PBC}}$ & Max$_{\text{PTBP}}$ & Max$_{\text{TB3PY}}$  \\
    \midrule
    \endfirsthead

    \caption*{Table \thetable{} - Continued from previous page} \\
    \toprule
    Mat. & JV\# & Spg. & Max$_{\text{matsci}}$ & Max$_{\text{PBC}}$ & Max$_{\text{PTBP}}$ & Max$_{\text{TB3PY}}$  \\
    \midrule
    \endhead

    \midrule
    \multicolumn{7}{r}{\textit{Continued on next page}} \\
    \endfoot

    \bottomrule
    \endlastfoot

    Cu & 867 & Fm$\bar{3}$m & 3.94 & - & 3.78 & 3.74  \\  
        Al & 816 & Fm$\bar{3}$m & 3.51 & - & 3.37 & 3.14  \\  
        Ti & 1029 & P6/mmm & - & - & 1.67 & 0.4  \\  
        Au & 825 & Fm$\bar{3}$m & - & - & 3.55 & 4.16  \\  
        GaAs & 1174 & F$\bar{4}$3m & - & - & 4.13 & 1.3  \\  
        Si & 1002 & Fd$\bar{3}$m & 5.43 & 4.07 & 4.63 & 4.54  \\  
        ZnO & 1195 & P6$_3$mc & - & - & 3.78 & 1.79  \\  
        SiC & 8118 & P6$_3$mc & 5.56 & 5.3 & 5.53 & 6.58  \\  
        SiC & 8158 &  F$\bar{4}$3m  & 5.85 & 5.47 & 4.79 & 5.37  \\  
        SiC & 107 & P6$_3$mc & 4.63 & 5.16 & 4.78 & 5.54  \\  
        AlP & 1327 & F$\bar{4}$3m & 5.94 & - & 5.36 & 3.95  \\  
        C & 91 & Fd$\bar{3}$m & 6.95 & 7.2 & 7.29 & 5.98  \\  
        SiO$_2$ & 41 & P3$_2$21 & 7.52 & 7.89 & 6.34 & 6.24  \\  
        SiO$_2$ & 34674 & C222$_1$ & 7.34 & 8.05 & 6.77 & 6.61  \\  
        TiO$_2$ & 104 & I4$_1$/amd & 3.47 & - & 2.61 & 0.64  \\  
        ZrO$_2$ & 113 & P2$_1$/c & - & - & 4.34 & -  \\  
        KCl & 1145 & Fm$\bar{3}$m & - & - & 5.01 & 5.68  \\  
        MgO & 116 & Fm$\bar{3}$m & - & - & 5.36 & 6.45  \\  
        InN & 1180 & P6$_3$mc & - & - & 3.38 & 2.63  \\  
        InP & 1183 & F$\bar{4}$3m & - & - & 4.87 & 0.52  \\  
        InSb & 1189 & F$\bar{4}$3m & - & - & 4.08 & 0.79  \\  
        ZnTe & 1198 & F$\bar{4}$3m & - & - & 6.04 & 4.37  \\  
        CuCl & 1201 & F$\bar{4}$3m & - & - & 2.68 & 0.85  \\  
        Cu$_2$O & 1216 & Pn$\bar{3}$m & 2.85 & - & - & -  \\  
        BaTe & 1267 & Fm$\bar{3}$m & - & - & 4.54 & 4.81  \\  
        BaSe & 1294 & Fm$\bar{3}$m & - & - & 3.95 & 5.14  \\  
        MgS & 1300 & Fm$\bar{3}$m & - & - & 4.86 & 5.43  \\  
        BP & 1312 & F$\bar{4}$3m & - & - & 5.42 & 5.34  \\  
        BaS & 1315 & Fm$\bar{3}$m & - & - & 3.74 & 2.07  \\  
        GaP & 1393 & F$\bar{4}$3m & - & - & 5.51 & 5.11  \\  
        AlSb & 1408 & F$\bar{4}$3m & - & - & 4.96 & 1.46  \\  
        AlCuO$_2$ & 1453 & R$\bar{3}$m & 4.04 & - & 3.83 & 1.54  \\  
        BN & 17 & P6$_3$/mmc & 5.07 & - & 4.99 & 8.39  \\  
        ZnS & 1702 & F$\bar{4}$3m & - & - & 5.46 & 0.63  \\  
        AgCl & 1954 & Fm$\bar{3}$m & - & - & 4.9 & 4.12  \\  
        CdTe & 23 & F$\bar{4}$3m & - & - & 4.88 & 3.4  \\  
        SnSe & 299 & Pnma & - & - & - & 0.8  \\  
        GaN & 30 & P6$_3$mc & - & - & 3.47 & 5.46  \\  
        Al$_2$O$_3$ & 32 & R$\bar{3}$c & 5.52 & - & 3.42 & -  \\  
        AlN & 39 & P6$_3$mc & - & - & 4.74 & 6.33  \\  
        TiO$_2$ & 5 & P4$_2$/mnm & 1.08 & - & 1.6 & 1.3  \\  
        MoS$_2$ & 54 & P6$_3$/mmc & - & - & 1.48 & 1.17  \\  
        MoSe$_2$ & 57 & P6$_3$/mmc & - & - & 1.89 & 1.18  \\  
        BAs & 7630 & F$\bar{4}$3m & - & - & 5.13 & 4.75  \\  
        MgSe & 7678 & Fm$\bar{3}$m & - & - & 5.09 & 6.04  \\  
        MgTe & 7762 & F$\bar{4}$3m & - & - & 4.21 & 4.53  \\  
        AlN & 7844 & F$\bar{4}$3m & - & - & 5.41 & 5.6  \\  
        SnTe & 7860 & Fm$\bar{3}$m & - & - & 3.94 & 0.32  \\  
        CdS & 8003 & F$\bar{4}$3m & - & - & 4.55 & 4.29  \\  
        GaN & 8169 & F$\bar{4}$3m & - & - & 4.49 & 5.52  \\  
        AgI & 8566 & Fm$\bar{3}$m & - & - & 3.56 & 3.19  \\  
        AgBr & 8583 & Fm$\bar{3}$m & - & - & 3.76 & 3.13  \\  
        Ge & 890 & Fd$\bar{3}$m & - & - & 3.39 & 1.61  \\  
        CdS & 95 & P6$_3$mc & - & - & 3.95 & 4.22  \\  
        ZnSe & 96 & F$\bar{4}$3m & - & - & 6.51 & 3.92  \\  
        InAs & 97 & F$\bar{4}$3m & - & - & 4 & 3.69  \\  
        Ni & 943 & Fm$\bar{3}$m & - & - & 3.7 & 1.18  \\  
        Ag & 813 & Fm$\bar{3}$m & - & - & 0.55 & 1.31  \\  
        MgB$_2$ & 1151 & P6/mmm & - & - & 3.57 & 3.8  \\  
        Nb & 934 & Im$\bar{3}$m & - & - & 3.7 & 3.69 \\
        \midrule
        Average & - & - & 4.92 & 6.16 & 4.26 & 3.61 \\

\end{longtable}
\end{landscape}

To further investigate structural dependence of electronic accuracy, we computed $\text{Diff}_{\text{max}}$ across \textcolor{black}{50+} materials, as shown in Table~\ref{maxdiff_comparison}. This metric captures the largest deviation in band eigenvalues between TB and OPT-DFT for each material. The average $\text{Diff}_{\text{max}}$ for DFTB-PTBP is 4.26~eV, while TB3PY yields a lower average value of 3.61~eV. \textcolor{black}{These results are consistent with the bandgap analysis and further confirm the superior electronic structure fidelity of the TB3PY model in most cases, especially in accurately tracking the eigenvalue spectra, which benefits from its angular-dependent terms.} 

To complement the electronic structure evaluation, we also assessed the predictive capabilities of the TB models in estimating mechanical properties, specifically the bulk modulus. This quantity reflects a material's resistance to uniform compression and is a critical mechanical parameter for electronic device applications, particularly in environments where thermal and mechanical stresses are prevalent.  Table~\ref{bulk_modulus_comparison} summarizes the bulk modulus values (in GPa) predicted by the various TB models and compares them with \textcolor{black}{OPT-DFT} reference values from the JARVIS database. \textcolor{black}{To better contextualize the performance and account for the varying magnitudes of bulk moduli, we report both absolute errors and Relative Errors (RE), where RE for a given material is calculated as the absolute difference between the predicted and reference values divided by the reference value. Subsequently, we also calculate the Mean Absolute Relative Error (MARE) for bulk moduli, representing the average of these individual relative errors across the entire test set. The materials in the table span a range of metallic, semiconducting, and insulating systems, offering insight into the transferability and physical fidelity of the models.}

\textcolor{black}{The bulk moduli results indicate that both TB models, while capturing general trends, exhibit significant quantitative deviations from OPT-DFT reference values, with Mean Absolute Relative Errors (MARE) of 79.2 \% for PTBP-DFTB and 102.3 \% for TB3PY. These large relative errors are partly a scale issue, as bulk moduli values span a wide range, and partly reflect that the current parameterizations of these models are primarily optimized for electronic structure (e.g., band energies and occupations) rather than highly accurate mechanical properties. The transferability challenges for bulk moduli are particularly pronounced in systems like SiC and AlP, where the errors are consistently high. These findings emphasize that the careful selection of TB models, and their specific parameterizations, must be dependent not only on material type but also on the specific properties being investigated.}

\begin{landscape}
\footnotesize
\setlength{\tabcolsep}{2.5pt}
\setlength{\LTcapwidth}{0.85\textwidth}
\begin{longtable}{@{}l>{\centering\arraybackslash}p{0.75cm}>{\centering\arraybackslash}p{1.15cm}*{4}{>{\centering\arraybackslash}p{0.9cm}}*{3}{>{\centering\arraybackslash}p{1cm}}@{}}
    \caption{Comparison of bulk modulus ($K$, in GPa) from TB methods with OPT-DFT (JARVIS). Relative Error: RE $= |K_{\text{TB}} - K_{\text{OPT}}| / K_{\text{OPT}} \times 100\%$. MARE is the mean of RE across materials.}
    \label{bulk_modulus_comparison} \\
    \toprule
    Mat. & JV\# & Spg. & $K_{\text{mat}}$ & $K_{\text{PBC}}$ & $K_{\text{PTB}}$ & $K_{\text{TB3}}$ & $K_{\text{OPT}}$ & RE$_{\text{PTB}}$ & RE$_{\text{TB3}}$ \\
    \midrule
    \endfirsthead

    \caption*{Table \thetable{} - Continued from previous page} \\
    \toprule
    Mat. & JV\# & Spg. & $K_{\text{mat}}$ & $K_{\text{PBC}}$ & $K_{\text{PTB}}$ & $K_{\text{TB3}}$ & $K_{\text{OPT}}$ & RE$_{\text{PTB}}$ & RE$_{\text{TB3}}$ \\
    \midrule
    \endhead

    \midrule
    \multicolumn{10}{r@{}}{\textit{Continued on next page}} \\
    \endfoot

    \bottomrule
    \endlastfoot

    Cu & 867 & Fm$\bar{3}$m & 68.81 & - & 42.35 & 34.49 & 141.4 & 70.05 & 75.61  \\  
    Al & 816 & Fm$\bar{3}$m & 34.74 & - & 31.05 & 16.53 & 69.93 & 55.60 & 76.36  \\  
    Ti & 1029 & P6/mmm & - & - & 24.68 & 18.66 & 115.17 & 78.57 & 83.80  \\  
    Au & 825 & Fm$\bar{3}$m & - & - & 42.2 & 28.93 & 148.6 & 71.60 & 80.53  \\  
    GaAs & 1174 & F$\bar{4}$3m & - & - & 22.45 & 13.92 & 61.93 & 63.75 & 77.52  \\  
    Si & 1002 & Fd$\bar{3}$m & 89.41 & 16.61 & 14.42 & 10.29 & 87.27 & 83.48 & 88.21  \\  
    ZnO & 1195 & P6$_3$mc & - & - & 29.91 & 26.84 & 137.33 & 78.22 & 80.46  \\  
    SiC & 8118 & P6$_3$mc & 36.71 & 48.11 & 31.99 & 26.94 & 213.53 & 85.02 & 87.38  \\  
    SiC & 8158 & F$\bar{4}$3m & 33.03 & 1 & 29.04 & 26.21 & 212.77 & 86.35 & 87.68  \\  
    SiC & 107 & P6$_3$mc & 36.68 & 47.02 & 31.17 & 26.75 & 213.34 & 85.39 & 87.46  \\  
    AlP & 1327 & F$\bar{4}$3m & 15.12 & - & 19.93 & 12.02 & 83.37 & 76.09 & 85.58  \\  
    C & 91 & Fd$\bar{3}$m & 81.05 & 88.18 & 75.18 & 70.13 & 437.4 & 82.81 & 83.97  \\  
    SiO$_2$ & 41 & P3$_2$21 & 8.75 & 11.44 & 8.18 & 22.39 & 39.41 & 79.24 & 43.19  \\  
    SiO$_2$ & 34674 & C222$_1$ & 13.34 & 5.39 & 4.35 & 22.42 & 41.54 & 89.53 & 46.03  \\  
    TiO$_2$ & 104 & I4$_1$/amd & 37.15 & - & 27.25 & 164.03 & 196.6 & 86.14 & 16.57  \\  
    ZrO$_2$ & 113 & P2$_1$/c & - & - & 33.45 & 100.43 & 187.44 & 82.15 & 46.42  \\  
    KCl & 1145 & Fm$\bar{3}$m & - & - & 3.68 & 3.63 & 20.33 & 81.90 & 82.14  \\  
    MgO & 116 & Fm$\bar{3}$m & - & - & 28.3 & 29.06 & 160.67 & 82.39 & 81.91  \\  
    InN & 1180 & P6$_3$mc & - & - & 26.02 & 20.8 & 126.27 & 79.39 & 83.53  \\  
    InP & 1183 & F$\bar{4}$3m & - & - & 19.07 & 9.8 & 59.7 & 68.06 & 83.58  \\  
    InSb & 1189 & F$\bar{4}$3m & - & - & 13.58 & 4.75 & 38.13 & 64.38 & 87.54  \\  
    ZnTe & 1198 & F$\bar{4}$3m & - & - & 15.31 & 7.77 & 45.33 & 66.23 & 82.86  \\  
    CuCl & 1201 & F$\bar{4}$3m & - & - & 1.05 & 9.29 & 53.27 & 98.03 & 82.56  \\  
    Cu$_2$O & 1216 & Pn$\bar{3}$m & 34.31 & - & - & 350.59 & 115.17 & - & 204.41  \\  
    BaTe & 1267 & Fm$\bar{3}$m & - & - & 8.48 & 5.4 & 31.47 & 73.05 & 82.84  \\  
    BaSe & 1294 & Fm$\bar{3}$m & - & - & 10.54 & 5.51 & 39.63 & 73.40 & 86.10  \\  
    MgS & 1300 & Fm$\bar{3}$m & - & - & 9.34 & 16.5 & 78.3 & 88.07 & 78.93  \\  
    BP & 1312 & F$\bar{4}$3m & - & - & 33.62 & 37.27 & 161.63 & 79.20 & 76.94  \\  
    BaS & 1315 & Fm$\bar{3}$m & - & - & 9.85 & 6.92 & 45.4 & 78.30 & 84.76  \\  
    GaP & 1393 & F$\bar{4}$3m & - & - & 28.18 & 8.16 & 77.43 & 63.61 & 89.46  \\  
    AlSb & 1408 & F$\bar{4}$3m & - & - & 14.32 & 7.61 & 50.43 & 71.60 & 84.91  \\  
    AlCuO$_2$ & 1453 & R$\bar{3}$m & 93.45 & - & 30.05 & 1754.5 & 181.81 & 83.47 & 865.02  \\  
    BN & 17 & P6$_3$/mmc & 81.29 & - & 33.7 & 41.4 & 245.04 & 86.25 & 83.10  \\  
    ZnS & 1702 & F$\bar{4}$3m & - & - & 13.9 & 12.13 & 72.53 & 80.84 & 83.28  \\  
    AgCl & 1954 & Fm$\bar{3}$m & - & - & 8.48 & 8.65 & 50 & 83.04 & 82.70  \\  
    CdTe & 23 & F$\bar{4}$3m & - & - & 10.54 & 5.8 & 37.83 & 72.14 & 84.67  \\  
    SnSe & 299 & Pnma & - & - & - & 8.34 & 29.74 & - & 71.96  \\  
    GaN & 30 & P6$_3$mc & - & - & 34.38 & 1080.02 & 178.97 & 80.79 & 503.46  \\  
    Al$_2$O$_3$ & 32 & R$\bar{3}$c & 141.64 & - & 24.57 & 731.43 & 241.28 & 89.82 & 203.15  \\  
    AlN & 39 & P6$_3$mc & - & - & 18.96 & 24.19 & 199.4 & 90.49 & 87.87  \\  
    TiO$_2$ & 5 & P4$_2$/mnm & 49.24 & - & 36.63 & 49.69 & 226.3 & 83.81 & 78.04  \\  
    MoS$_2$ & 54 & P6$_3$/mmc & - & - & 17.42 & 32.91 & 70.51 & 75.29 & 53.33  \\  
    MoSe$_2$ & 57 & P6$_3$/mmc & - & - & 11.32 & 24.1 & 57.63 & 80.36 & 58.18  \\  
    BAs & 7630 & F$\bar{4}$3m & - & - & 28.29 & 21.48 & 133.37 & 78.79 & 83.89  \\  
    MgSe & 7678 & Fm$\bar{3}$m & - & - & 10.57 & 10.51 & 65 & 83.74 & 83.83  \\  
    MgTe & 7762 & F$\bar{4}$3m & - & - & 8.71 & 4.59 & 36.67 & 76.25 & 87.48  \\  
    AlN & 7844 & F$\bar{4}$3m & - & - & 1 & 25.68 & 198.63 & 99.50 & 87.07  \\  
    SnTe & 7860 & Fm$\bar{3}$m & - & - & 11.66 & 6.21 & 42.1 & 72.30 & 85.25  \\  
    CdS & 8003 & F$\bar{4}$3m & - & - & 11.02 & 11.41 & 57.1 & 80.70 & 80.02  \\  
    GaN & 8169 & F$\bar{4}$3m & - & - & 26.41 & 33.65 & 178.67 & 85.22 & 81.17  \\  
    AgI & 8566 & Fm$\bar{3}$m & - & - & 1 & 6.76 & 40.27 & 97.52 & 83.21  \\  
    AgBr & 8583 & Fm$\bar{3}$m & - & - & 7.26 & 8.31 & 46.07 & 84.24 & 81.96  \\  
    Ge & 890 & Fd$\bar{3}$m & - & - & 16.34 & 8.09 & 58.07 & 71.86 & 86.07  \\  
    CdS & 95 & P6$_3$mc & - & - & 14.22 & 10.98 & 57.03 & 75.07 & 80.75  \\  
    ZnSe & 96 & F$\bar{4}$3m & - & - & 17 & 10 & 59.7 & 71.52 & 83.25  \\  
    InAs & 97 & F$\bar{4}$3m & - & - & 17.98 & 9.28 & 50.03 & 64.06 & 81.45  \\  
    Ni & 943 & Fm$\bar{3}$m & - & - & 34.48 & 40.46 & 200.43 & 82.80 & 79.81  \\  
    Ag & 813 & Fm$\bar{3}$m & - & - & 24.02 & 22.44 & 100.27 & 76.04 & 77.62  \\  
    MgB$_2$ & 1151 & P6/mmm & - & - & 25.36 & 211.73 & 155.42 & 83.68 & 36.23  \\  
    Nb & 934 & Im$\bar{3}$m & - & - & 29.33 & 26.73 & 175.67 & 83.30 & 84.78 \\
    \midrule
    MAE & - & - & 118.48 & 146.79 & 92.71 & 131.02 & - & - & - \\
    \midrule
    MARE & - & - & - & - & - & - & - & 79.22\% & 102.30\% \\

\end{longtable}
\end{landscape}

\section{Conclusion}

\textcolor{black}{In summary, CHIPS-TB presents a streamlined computational benchmarking framework for evaluating tight-binding models using high-quality OPT-DFT reference data from the JARVIS database. As a complimentary to the machine learning force-field benchmarking effort, (e.g., CHIPS-FF) it focuses on the unique challenges and insights associated with electronic structure properties (bandgaps, band dispersion, etc.) rather than purely force-field or structural properties. Our systematic comparisons, informed by a detailed understanding of the underlying formalisms of DFTB, SlaKoNet and TB3PY, provide critical insights into their respective strengths and limitations.}

\textcolor{black}{For bandgap prediction across 50+ materials, TB3PY achieved a lower MAE of 0.67~eV against OPT-DFT, demonstrating superior fidelity compared to DFTB-PTBP's MAE of 0.82~eV. The SlaKoNet model showed a remarkable MAE of 0.81~eV against experimental bandgaps and 0.76~eV against mBJ-DFT, confirming the benefits of training on higher-accuracy references.
When considering the overall eigenvalue distributions ($\text{Diff}_{\text{max}}$), TB3PY exhibited a lower average maximum deviation (3.61~eV) compared to DFTB (4.26~eV), particularly in reproducing eigenvalue spectra in metallic systems like aluminum. This suggests that the angular-dependent three-body terms in TB3PY contribute to a more accurate description of band curvature, which is essential for capturing delocalized interactions in metals. Conversely, DFTB's reliance on approximate Kohn-Sham DFT and two-center integral tabulations, despite its self-consistent charge corrections, may lead to flatter bands and overestimation of bandwidths in certain metallic contexts. For bulk moduli, both models show significant absolute and relative errors, highlighting a current limitation in their transferability to mechanical properties, likely due to parameterization priorities.}

\textcolor{black}{These findings underscore that while TB3PY generally provides improved accuracy for eigenvalue distributions, DFTB's PTBP parameterization performs favorably for bandgap prediction in this specific test set. The results emphasize the crucial need for careful selection of TB models and their specific parameterizations, dependent on both material type and the targeted properties. These insights also point to opportunities for improving parameterization strategies, such as data-driven fitting or hybrid methods, to enhance the accuracy and transferability of TB models in electronic structure simulations. Future work will expand this framework to study energetics, defect, interface, and transport properties of semiconductor devices, incorporating a larger class and set of materials, potentially including data from the JARVIS-DFT, and exploring the integration of advanced TB variants like LC-DFTB.}

\section*{Acknowledgments}

All authors thank the National Institute of Standards and Technology for funding, computational, and data-management resources. K.C. acknowledges computational resources from the Johns Hopkins University. This work was performed with funding from the CHIPS Metrology Program, part of CHIPS for America, National Institute of Standards and Technology, U.S. Department of Commerce. Certain commercial equipment, instruments, software, or materials are identified in this paper in order to specify the experimental procedure adequately. Such identifications are not intended to imply recommendation or endorsement by NIST, nor it is intended to imply that the materials or equipment identified are necessarily the best available for the purpose. The authors would like to acknowledge Kevin Garrity (NIST) for fruitful discussion.

%
%



 \bibliographystyle{elsarticle-num-names} 
\bibliography{bio}
\end{document}